\documentstyle[amstex,preprint,prb,aps]{revtex}

\begin{document}
\title{Wave Packet Spreading: Temperature and Squeezing Effects with
Applications to Quantum Measurement and Decoherence}
\author{G. W. Ford$^\dag$ and R. F. O'Connell$^\ddag$}
\address{School of Theoretical Physics\\
Dublin Institute for Advanced Studies\\
10 Burlington Road, Dublin 4, Ireland}
\maketitle

\begin{abstract}
A localized free particle is represented by a wave packet and its motion is
discussed in most quantum mechanics textbooks. Implicit in these discussions
is the assumption of zero temperature. We discuss how the effects of finite
temperature and squeezing can be incorporated in an elementary manner. The
results show how the introduction of simple tools and ideas can bring the
reader into contact with topics at the frontiers of research in quantum
mechanics. We discuss the standard quantum limit, which is of interest in
the measurement of small forces, and decoherence of a mixed (``Schr\"{o}%
dinger cat'') state, which has implications for current research in quantum
computation, entanglement, and the quantum-classical interface.
\end{abstract}


\section{Introduction}

The uncertainty principle in quantum mechanics implies that the position and
momentum of a particle cannot be determined simultaneously with arbitrary
precision. More explicitly
\begin{equation}
\Delta {x}\Delta {p}\geq \frac{\hbar }{2},  \label{1.1}
\end{equation}%
where $(\Delta x)^{2}=\langle (x-\langle {x}\rangle )^{2}\rangle $ and $%
(\Delta p)^{2}=\langle {(}p-\langle {p}\rangle {)}^{2}\rangle $ are the
variance of the position and momentum of the particle, respectively. For a
noninteracting particle the expected value at time $t$ of an operator ${\cal %
O}$ is given by
\begin{equation}
\langle {\cal O}(t)\rangle =\int_{-\infty }^{\infty }dx\,\psi ^{\ast }(x,t)%
{\cal O}\psi {(}x,t),  \label{1.3}
\end{equation}%
where the wave function $\psi {(}x,t)$ is the solution of the time-dependent
Schr\"{o}dinger equation,
\begin{equation}
i\hbar \frac{\partial \psi }{\partial t}=H\psi ,  \label{1.4}
\end{equation}%
with $H$ the particle Hamiltonian. For the more general case, such as a
particle interacting with a heat bath, the expected value would be given by %
\cite{fano}
\begin{equation}
\langle {\cal O}(t)\rangle ={\rm Tr}\{\rho (t){\cal O}\},  \label{1.5}
\end{equation}%
where the density matrix $\rho (t)$ is the solution of the von Neuman
equation,
\begin{equation}
i\hbar \frac{\partial \rho }{\partial t}=[H,\rho ],  \label{1.6}
\end{equation}%
with $H$ now the Hamiltonian for the entire system.

For a free particle, the stationary solutions of the Schr\"{o}dinger
equation are plane waves, for which the particle may be found with equal
probability anywhere in space, that is, $\Delta x=\infty$, $\Delta p=0$.
However, one is often interested in describing a localized particle, which
can be achieved by constructing a wave packet corresponding to a
superposition of plane waves. Such a packet is necessarily not stationary
and will spread (or shrink) in time. Wave packet spreading is of fundamental
interest, appears in many contexts, and is discussed in introductory\cite%
{liboff} and advanced\cite{schiff,merzbacher} quantum mechanics
textbooks.

We begin our discussion in Sec.~II, where we describe the motion of an
arbitrary free-particle wave packet. This description is more or less
standard, the main result being a general expression for the width at time $%
t $ in term of the initial data. This expression does not, however, take
into account the uncertainty principle; to do so one must evaluate the
initial data using Eqs.~(\ref{1.3}) or (\ref{1.5}). We do this in Sec.~III,
where we begin with a brief derivation of the uncertainty principle. There
we introduce annihilation and creation operators that are analogous to those
appearing in discussions of the harmonic oscillator,\cite%
{liboff,schiff,merzbacher} but which now apply to an arbitrary state. An
immediate consequence is a simple construction of the most general minimum
uncertainty wave packet. We conclude Sec.~III with an expression for the
spreading of an initial minimum uncertainty state.

In Sec.~IV we consider the effect of finite temperature on wave packet
spreading. An initial wave packet at finite temperature is in a mixed state:
there is no corresponding wave function, and the state is described by a
density matrix. Nevertheless, we can calculate the temperature effect by
forming the observed quantities (the probability distribution or the
expected values) with an initial wave function and then averaging over a
Maxwell distribution of initial velocities. The result is an additional
spreading that dominates when the thermal de Broglie wavelength is small
compared to the initial width.

In Sec.~V we consider spreading for squeezed states. Squeezing is generally
discussed in the context of the harmonic oscillator, but here we discuss
squeezing of a Gaussian wave packet. An interesting result is that, while
for sufficiently long times the root-mean square (rms) width of such a wave
packet increases linearly with time, for short times it can even shrink.
Finally, in Sec.~VI we consider two topical applications: the standard
quantum limit arising in connection with the measurement of small forces and
decoherence. In particular, we give a new and simple demonstration of how to
circumvent the standard quantum limit. This limit is of interest not only
for gravitational wave detection, but for any application where the question
of the accuracy of successive measurements arises. With regard to the
decoherence problem, the temperature effect on wave packet
spreading is an essential feature.

\section{Free Particle}

We begin by reminding ourselves that from either Eq.~(\ref{1.3}) or (\ref%
{1.5}), we can show that the rate of change of the expected value of an
operator ${\cal O}$, with no explicit time dependence, is given
by:
\begin{equation}
i\hbar {\frac{d\left\langle {\cal O}\right\rangle }{dt}}=\left\langle [{\cal %
O},H]\right\rangle .  \label{2.1}
\end{equation}%
For a free particle with $H=p^{2}/2m$, we can use the canonical commutation
relation,
\begin{equation}
\lbrack x,p]=i\hbar ,  \label{2.3}
\end{equation}%
to show that
\begin{equation}
{\frac{d\left\langle x\right\rangle }{dt}}={\frac{\left\langle
p\right\rangle }{m}},\qquad {\frac{d\left\langle p\right\rangle }{dt}}=0.
\label{2.4}
\end{equation}%
These are the classical equations of motion (Ehrenfest theorem). Therefore,
we have
\begin{equation}
\left\langle x(t)\right\rangle =\left\langle x(0)\right\rangle +{\frac{%
\left\langle p(0)\right\rangle }{m}}t,\qquad \left\langle p(t)\right\rangle
=\left\langle p(0)\right\rangle .  \label{2.5}
\end{equation}%
In the same way, we can show that
\begin{equation}
{\frac{d\left\langle x^{2}\right\rangle }{dt}}={\frac{\left\langle
xp+px\right\rangle }{m}},\quad {\frac{d\left\langle xp+px\right\rangle }{dt}}%
={\frac{2\left\langle p^{2}\right\rangle }{m}},\quad {\frac{d\left\langle
p^{2}\right\rangle }{dt}}=0,  \label{2.6}
\end{equation}%
and therefore,
\begin{equation}
\left\langle x^{2}(t)\right\rangle =\left\langle x^{2}(0)\right\rangle +{%
\frac{\left\langle x(0)p(0)+p(0)x(0)\right\rangle }{m}}t+{\frac{\left\langle
p^{2}(0)\right\rangle }{m^{2}}}t^{2}.  \label{2.7}
\end{equation}%
We can write the above results in terms of the variances as
\begin{equation}
\Delta {x}^{2}(t)=\Delta {x}^{2}(0)+\frac{1}{m^{2}}\Delta {p}^{2}(0)t^{2}+%
\frac{\left\langle {x}(0){p}(0)+{p}(0)x(0)\right\rangle -2\left\langle
x(0)\right\rangle \left\langle p(0)\right\rangle }{m}t.  \label{2.8}
\end{equation}%
Hence, for sufficiently long times, $\Delta x(t)$, the rms width of the wave
packet at time $t$, increases linearly with time. However, it is possible
for a wave packet to shrink for a time, as we shall discuss in Sec.~V.

In the formal solution of the equations of mean motion, it appears that the
initial data, $\Delta {x}^{2}(0)$, $\Delta {p}^{2}(0)$ and $\left\langle {x}%
(0){p}(0)+{p}(0)x(0)\right\rangle -2\left\langle x(0)\right\rangle
\left\langle p(0)\right\rangle$, could be given arbitrary values. We
emphasize that this is not so, and the initial expectations must be obtained
from the initial state by an expression of the form (\ref{1.3}) or (\ref{1.5}%
). In particular, the initial variances must satisfy the uncertainty
principle (\ref{1.1}).

\section{Minimal (non-squeezed) Gaussian Wave Packet}

To make the discussion as simple as possible, we begin by restricting our
discussion to states for which $\left\langle x\right\rangle=0$ and $%
\left\langle p\right\rangle=0$. We then introduce the operators
\begin{equation}
a=\frac{x}{2\sigma}+i\frac{\sigma p}{\hbar},\qquad a^{\dag}=\frac{x}{2\sigma}%
-i\frac{\sigma p}{\hbar},  \label{3.1}
\end{equation}
where $\sigma$ is a real parameter. These operators are formally identical
with the annihilation and creation operators usually introduced in
connection with the harmonic oscillator,\cite{liboff,schiff,merzbacher} but
here they apply to an arbitrary state (pure or mixed) without reference to
an external potential. Next, we form the necessarily positive quantity,
\begin{equation}
\left\langle a^{\dag}a\right\rangle =\frac{\Delta x^{2}}{4\sigma^{2}}+\frac{%
\sigma^{2}\Delta p^{2}}{\hbar^{2}}-\frac{1}{2}\geq 0,  \label{3.2}
\end{equation}
where we have used the canonical commutation relation in Eq.~(\ref{2.3}). We
seek the minimum of this quantity with respect to variations of $\sigma^{2}$%
, which occurs when
\begin{equation}
\sigma^{2}=\frac{\hbar \Delta x}{2\Delta p}.  \label{3.3}
\end{equation}
With this value of $\sigma^{2}$, we see that
\begin{equation}
\left\langle a^{\dag}a\right\rangle =\frac{\Delta x\Delta p}{\hbar}-\frac{1}{%
2}\geq 0,  \label{3.4}
\end{equation}
which is just the uncertainty principle. The minimum uncertainty state, for
which the inequality becomes an equality, must be a pure state that
corresponds to a wave function $\phi$ satisfying $\left\langle
a^{\dag}a\right\rangle =\left\| a\phi \right\|^{2}=0$. That is, $\phi$ must
satisfy
\begin{equation}
a\phi =(\frac{x}{2\sigma}+\sigma \frac{d}{dx})\phi =0,  \label{3.5}
\end{equation}
where we have used the familiar realization of the momentum operator: $p=%
\frac{\hbar}{i}\frac{d}{dx}$. The solution of the first-order differential
equation in Eq.~(\ref{3.5}) is
\begin{equation}
\phi(x)=\frac{1}{(2\pi \sigma^{2})^{1/4}} e^{-x^2/4\sigma^2},  \label{3.6}
\end{equation}
where we have chosen the normalization so that $\int_{-\infty}^{\infty}dx\,%
\phi^{\ast}(x)\phi (x)=1$.

To extend this result for nonvanishing $\left\langle x\right\rangle $ and $%
\left\langle p\right\rangle $, we need only make the replacements $%
x\rightarrow x-\left\langle x\right\rangle $ and $p\rightarrow
p-\left\langle p\right\rangle $ in Eq.~(\ref{3.1}) and repeat the argument.
The result is that the most general minimum uncertainty wave packet has the
form:
\begin{equation}
\phi (x)=\frac{1}{(2\pi \sigma ^{2})^{1/4}}\exp \left\{ -\frac{(x-x_{0})^{2}%
}{4\sigma ^{2}}+i\frac{mv_{0}x}{\hbar }\right\} ,  \label{3.8}
\end{equation}%
where $\sigma $, $x_{0}$ and $v_{0}$ are real. Thus the minimum uncertainty
wave packet is a Gaussian, centered at $x_{0}$ and moving with velocity $%
v_{0}$.

Suppose we choose the initial state to be a minimum uncertainty state with $%
\psi (x,0)$ given by Eq.~(\ref{3.8}). Then we find,
\begin{gather}
\left\langle x(0)\right\rangle =x_{0},\qquad \left\langle p(0)\right\rangle
=mv_{0},  \nonumber \\
\left\langle x^{2}(0)\right\rangle =x_{0}^{2}+\sigma ^{2},\qquad
\left\langle p^{2}(0)\right\rangle =m^{2}v_{0}^{2}+{\frac{\hbar ^{2}}{%
4\sigma ^{2}}},  \nonumber \\
\left\langle x(0)p(0)+p(0)x(0)\right\rangle =2mx_{0}v_{0}.  \label{3.9}
\end{gather}%
With these expressions, Eq.~(\ref{2.8}) for the mean square width of the
wave packet becomes
\begin{equation}
\Delta x^{2}(t)=\sigma ^{2}+\left( {\frac{\hbar t}{2m\sigma }}\right) ^{2}.
\label{3.10}
\end{equation}%
The wave packet will expand so that the mean square width doubles in a time $%
t=2m\sigma ^{2}/\hbar $. During this time, the wave packet will have
traveled a distance $\ell =v_{0}t=4\pi \sigma ^{2}/\lambda $, where $\lambda
=mv_{0}/2\pi \hbar $ is the de Broglie wavelength.

Whereas our derivation might appear to be similar to that found
in some textbooks, there are important differences in that we
have allowed from the beginning the possibility of a mixed
state and we have shown that the minimum uncertainty state is a
Gaussian.  Furthermore, we have introduced the concept of
creation and annihilation operators for arbitrary states and,
concomitantly (as we shall see in Sec. V), this enables us to
consider squeezing of arbitrary states (as distinct from just
harmonic oscillator states).

\section{Effect of Temperature on the Spreading of a Free Particle Wave
Packet}

\label{sec:temp} We proceed by first calculating the wave function for the
particle at time $t$ and then forming the probability distribution. This
procedure is instructive, because it provides another method for calculating
the result in Eq.~(\ref{2.8}) for $\Delta {x}^{2}(t)$. Next, we take into
account the thermal distribution of initial velocities

Consider the solution of the free-particle Schr\"{o}dinger equation with a
given initial state. The general solution is\cite{merzbacher}
\begin{equation}
\psi (x,t)=\sqrt{\frac{m}{2\pi i\hbar t}}\int_{-\infty }^{\infty }dx^{\prime
}\exp \left\{ -\frac{m(x-x^{\prime })^{2}}{2i\hbar t}\right\} \psi
(x^{\prime },0).  \label{4.2}
\end{equation}%
We now apply this result to the case of an initial minimum uncertainty wave
packet with $\psi (x,0)$ of the general form given by Eq.~(\ref{3.8}). We
find,
\begin{equation}
\psi (x,t)=\frac{1}{[2\pi (\sigma +\frac{i\hbar t}{2m\sigma })^{2}]^{1/4}}%
\exp \left\{ -\frac{(x-x_{0}-v_{0}t)^{2}}{4\sigma ^{2}+\frac{2i\hbar t}{m}}+i%
\frac{mv_{0}}{\hbar }x-i\frac{mv_{0}^{2}t}{2\hbar }\right\} .  \label{4.3}
\end{equation}%
The probability distribution is
\begin{eqnarray}
P(x;t) &=&\left| \psi (x,t)\right| ^{2}  \nonumber \\
&=&\left[ 2\pi \Delta x^{2}(t)\right] ^{-1/2}\exp \left\{ -\frac{%
(x-x_{0}-v_{0}t)^{2}}{2\Delta x^{2}(t)}\right\} ,  \label{4.4}
\end{eqnarray}%
which is a Gaussian centered at the mean position of the particle at time $t$
with variance given by Eq.~(\ref{3.10}).

These results are standard quantum mechanics. Next we consider an ensemble
of particles in thermal equilibrium, but so weakly coupled to a heat bath
that we can neglect dissipation in the equation of motion. Each particle has
a wave function of the form (\ref{3.8}), with a Maxwell distribution of
initial velocities. (Note that the wave functions differ only by the phase
factor $\exp [imv_{0}x/\hbar ]$ and that the distribution in initial
velocities implies a corresponding distribution of the phase.) We obtain the
corresponding probability distribution by averaging the distribution (\ref%
{4.4}) over a thermal distribution of initial velocities. The result is
\begin{eqnarray}
P_{{\rm T}}(x;t) &=&\sqrt{\frac{m}{2\pi kT}}\int_{-\infty }^{\infty
}dv_{0}\!\exp \left\{ -\frac{mv_{0}^{2}}{2kT}\right\} P(x;t)  \nonumber \\
&=&\frac{1}{\sqrt{2\pi \Delta {x}_{{\rm T}}^{2}(t)}}\exp \left\{ -\frac{%
(x-x_{0})^{2}}{2\Delta {x}_{{\rm T}}^{2}(t)}\right\} ,  \label{4.5}
\end{eqnarray}%
in which%
\begin{eqnarray}
\Delta {x}_{{\rm T}}^{2}(t) &=&\Delta {x}^{2}(t)+{\frac{kT}{m}}t^{2}
\nonumber \\
&=&\sigma ^{2}+\left( \frac{\hbar ^{2}}{4m^{2}\sigma ^{2}}+\frac{kT}{m}%
\right) t^{2}.  \label{4.6}
\end{eqnarray}%
Here we have introduced the subscript $T$ to emphasize that the probability
distribution corresponds to finite temperature. Thus, there is an additional
spreading which is just that due to a Maxwell distribution of particle
velocities. As we shall see, in Sec. VI B, it is this extra term in the
spreading of the wave packet which is the origin of decoherence.

We could just as well obtain Eq.~(\ref{4.6}) by averaging the expressions in
Eq.~(\ref{3.9}) for the initial moments. The result is
\begin{gather}
\left\langle x(0)\right\rangle _{{\rm T}}=x_{0},\qquad \left\langle
p(0)\right\rangle _{{\rm T}}=0,  \nonumber \\
\left\langle x^{2}(0)\right\rangle _{{\rm T}}=x_{0}^{2}+\sigma ^{2},\qquad
\left\langle p^{2}(0)\right\rangle _{{\rm T}}=mkT+{\frac{\hbar ^{2}}{4\sigma
^{2}}},  \nonumber \\
\left\langle x(0)p(0)+p(0)x(0)\right\rangle _{{\rm T}}=0.  \label{4.7}
\end{gather}%
If we substitute these results into Eq.~(\ref{2.8}) for $\Delta x$, we find
the result (\ref{4.6}).

The center of the distribution $P_{T}(x,t)$ in Eq.~(\ref{4.5}) does not move
because the mean initial velocity is zero. The variance in Eq.~(\ref{4.6})
is the sum of three terms: the initial variance $\sigma ^{2}$, the
uncertainty principle spreading $(\hbar t/2m\sigma )^{2}$, and the thermal
spreading $\frac{kT}{m}t^{2}$. The ratio of the last two is
\begin{equation}
\frac{mkT\sigma ^{2}}{4\hbar ^{2}}=\frac{\sigma ^{2}}{4\bar{\lambda}^{2}},
\label{4.8}
\end{equation}%
where $\bar{\lambda}=\hbar /\sqrt{mkT}$ is the mean de Broglie wavelength.
Therefore, the thermal spreading will dominate when the mean de Broglie
wavelength is small compared to the initial width of the packet.

Note that the initial thermal state we have described is what is called a
mixed state. For such a state there is no single wave function with which
one can form observables such as the probability distribution in Eq.~(\ref%
{4.4}); rather the state is described by a density matrix.\cite{fano} We
have avoided introducing the density matrix to keep the discussion simple,
but for those who would like to see it, the matrix is given by
\begin{eqnarray}
\left\langle x\left| \rho (0)\right| x^{\prime }\right\rangle  &=&\overline{%
\phi (x)\phi ^{\ast }(x^{\prime })}  \nonumber \\
&=&\sqrt{\frac{m}{2\pi kT}}\!\int_{-\infty }^{\infty }dv_{0}\,\exp \left\{ -%
\frac{mv_{0}^{2}}{2kT}\right\} \phi \left( x\right) \phi ^{\ast }(x^{\prime
})  \nonumber \\
&=&\frac{1}{\sqrt{2\pi \sigma ^{2}}}\exp \left\{ -\frac{(x-x_{0})^{2}+(x^{%
\prime }-x_{0})^{2}}{4\sigma ^{2}}-\frac{mkT(x-x^{\prime })^{2}}{2\hbar ^{2}}%
\right\} ,  \label{4.10}
\end{eqnarray}%
where $\phi $ is given by Eq.~(\ref{3.8}). We also note that we have
neglected dissipation during the time development. Of course, in order to
come to thermal equilibrium, a particle must be coupled to a heat bath and
there must be a corresponding dissipation. The strength of this coupling
would be measured by a typical decay rate $\gamma $. If the coupling is
weak, we must wait a long time of order $\gamma ^{-1}$ for the system to
come to equilibrium, but the equilibrium state will be independent of
dissipation. The situation is like that for an ideal gas: collisions are
necessary to bring about an approach to equilibrium, but do not appear in
the equation of state nor in the velocity distribution. On the other hand,
the effect of dissipation on the time development of the initial state can
be neglected only for times short compared to $\gamma ^{-1}$. Our simple
expression (\ref{4.6}) for wave packet spreading is therefore valid only for
such short times ($\gamma t\ll 1$) where the motion is that of a free
particle; this short-time behavior is exactly what is relevant for the
calculation of decoherence times, as we shall discuss in detail in Sec.~\ref%
{sec:decohere}.

\section{Effects due to Squeezing}

In general, a squeezed state is defined as one in which the uncertainty of
one variable is reduced at the expense of an increase in its conjugate
variable.\cite{scully} If the original state is a minimum uncertainty state,
then the squeezed state may also be a minimum uncertainty state (the
so-called ideal squeezed state, such as a coherent state), but, more
generally it is not. We start with the simple case of a minimum uncertainty
state which is squeezed so that the uncertainty in $x$ remains unchanged but
the uncertainty in $p$ increases. The corresponding squeezed state is also
Gaussian,
\begin{equation}
\phi _{C}(x)=\frac{1}{(2\pi \sigma ^{2})^{1/4}}e^{-\frac{(1-iC)x^{2}}{%
4\sigma ^{2}}}.  \label{5.1}
\end{equation}%
This squeezed state can be represented as the result of a unitary operation
on the minimum uncertainty state (\ref{3.6}),
\begin{equation}
\phi _{C}(x)=e^{\frac{1}{4}iC(a+a^{\dag })^{2}}\phi (x).  \label{5.2}
\end{equation}%
If we use the general formula (\ref{1.3}) with $\psi (x,0)=\phi _{C}(x)$, we
find
\begin{gather}
\left\langle x(0)\right\rangle =0,\qquad \left\langle p(0)\right\rangle =0,
\nonumber \\
\left\langle x^{2}(0)\right\rangle =\sigma ^{2},\qquad \left\langle
p^{2}(0)\right\rangle =\frac{\hbar ^{2}(1+C^{2})}{4\sigma ^{2}},  \nonumber
\\
\left\langle x(0)p(0)+p(0)x(0)\right\rangle =\hbar C.  \label{5.3}
\end{gather}%
The squeezed state is therefore not a minimum uncertainty state, because $%
\Delta x\Delta p=\frac{\hbar }{2}\sqrt{1+C^{2}}>\frac{\hbar }{2}$.

With these expressions, Eq.~(\ref{2.8}) for the variance of the wave packet
becomes
\begin{equation}
\Delta {x}^{2}(t)=\sigma ^{2}\left( 1+{\frac{C\hbar {t}}{2\sigma ^{2}m}}%
\right) ^{2}+\left( {\frac{\hbar {t}}{2m\sigma }}\right) ^{2}.  \label{5.4}
\end{equation}%
If $C<0$, the wave packet first contracts, then expands; for very long times
the wave packet always expands.

Finally, it is clear from the analysis given in Sec.~\ref{sec:temp} that the
thermal contribution to the spreading is the same for both the squeezed and
unsqueezed states.

\section{Applications}

\subsection{Standard Quantum Limit}

Accurate measurements of the position of a free mass is a subject of much
current interest, in particular in the context of gravitational-wave
detection where questions have arisen as to whether there are fundamental
quantum mechanical limits on detection sensitivity. The point is that the
act of measurement introduces spreading which affects subsequent
measurements. In this context, Braginsky and Vorontsov\cite{brag} have
argued that in two successive measurements of the position $x$ of a free
mass $m$ made at a fixed time interval $t$, there is an uncertainty $\Delta {%
x(t)}$ in the result of the second measurement satisfying
\begin{equation}
\Delta {x(t)}\geq \sqrt{\frac{\hbar {t}}{m}}.  \label{6.1}
\end{equation}%
The inequality (\ref{6.1}) is called the standard quantum limit. On the
other hand, Yuen,\cite{yuen} while agreeing that this is the correct result
for free masses prepared in coherent states (i. e., minimum uncertainty
Gaussian states), has argued that the inequality can be violated for
squeezed states. We now give a simple derivation of the standard quantum
limit and Yuen's result.

Consider the squeezed state (\ref{5.1}) for which $\Delta {x}(0)=\sigma $
and $\Delta {x}^{2}(t)$ is given by Eq.~(\ref{5.4}). What is the value of $%
\sigma ^{2}$ for which $\Delta {x}^{2}(t)$ is a minimum for a given $t$? By
calculating the derivative with respect to $\sigma ^{2}$, we find that the
minimum occurs when
\begin{equation}
\sigma ^{2}=\sqrt{1+C^{2}}\frac{\hbar {t}}{2m}{.}  \label{6.2}
\end{equation}%
At the minimum, $\Delta {x}^{2}(t)$ has the value
\begin{equation}
\Delta {x}^{2}(t)_{{\rm min}}=(\sqrt{1+C^{2}}+C)\frac{\hbar {t}}{m}.
\label{6.3}
\end{equation}%
We therefore have in place of (\ref{6.1}) the general inequality
\begin{equation}
\Delta {x}(t)\geq \sqrt{(\sqrt{1+C^{2}}+C)\frac{\hbar {t}}{m}}.  \label{6.4}
\end{equation}%
For a minimum uncertainty state, where $C=0$, this corresponds to the
standard quantum limit (\ref{6.1}). On the other hand, for $C$ large and
negative, the right-hand side of (\ref{6.4}) can be as small as one likes,
as noted by Yuen.\cite{yuen}

While our explicit result (\ref{6.4}) supports Yuen's general conclusion
that more sensitive detection is possible when the initial measurement
results in a squeezed state, a more careful examination of the result makes
it clear that the may be difficult to achieve in practise. This is because
the condition (\ref{6.2}) puts a restriction on $\sigma $ ( the width of the
initial wave packet). In fact, a large negative $C$ implies that $\sigma $
be large, so the initial measurement must have large uncertainty. Moreover,
we see from (\ref{5.3}) and (\ref{6.2}) that, when the minimum value of $%
\Delta x(t)$ is achieved,
\begin{equation}
\langle {E}\rangle =\left\langle \frac{p^{2}}{2m}\right\rangle =\frac{%
m\sigma ^{2}}{2t^{2}}=\frac{\hbar }{4t}\sqrt{1+C^{2}}.  \label{3s8}
\end{equation}%
Thus the energy needed to produce the state increases with increasing $C$.

\subsection{Decoherence}

\label{sec:decohere} Decoherence refers to the destruction of a quantum
interference pattern and is relevant to the many experiments that depend on
achieving and maintaining entangled states. Examples of such efforts are in
the areas of quantum teleportation,\cite{zeilinger1} quantum information and
computation,\cite{bennett,quantum} entangled states,\cite{haroche} Schr\"{o}%
dinger cats,\cite{zeilinger2} and the quantum-classical interface.\cite%
{tegmark} For an overview of many of the interesting experiments involving
decoherence, we refer to Refs.~\onlinecite{haroche} and \onlinecite{myatt}.

Much of the discussion of decoherence\cite{giulini,ford1,ford2,ford3} has
been in terms of a particle moving in one dimension that is placed in an
initial superposition state (Schr\"{o}dinger ``cat'' state) corresponding to
two widely separated wave packets, each of the form (\ref{3.8}) but having $%
x_{0}=\pm \frac{d}{2}$ so that the packages are separated by a distance $d$.
Thus, in an obvious notation we write the wave-function of the two-Gaussian
state as
\begin{equation}
\psi (x,t)=N\left[ \psi _{1}(x,t)+\psi _{2}(x,t)\right] ,  \label{3s9}
\end{equation}%
where $\psi _{1}$ and $\psi _{2}$ are each given by the right-side of Eq. (%
\ref{4.3}), but with $x_{0}$ replaced by $\frac{d}{2}$ and $-\frac{d}{2}$,
respectively, and the normalization constant is%
\begin{equation}
N=\frac{1}{\sqrt{2(1+e^{-d^{2}/8\sigma ^{2}})}}.  \label{3s9a}
\end{equation}%
Hence
\begin{equation}
P(x,t)=N^{2}(|\psi _{1}|^{2}+|\psi _{2}|^{2}+2{\rm Re}\{\psi _{1}^{\ast
}\psi _{2}\}).  \label{4s0}
\end{equation}%
Thus, the probability distribution consists of three contributions, two of
which correspond to the separate packets, whereas the third is an
interference term. For a free particle at rest at zero temperature (i. e.,
in (\ref{4.3})we take $v_{0}=0$), an elementary calculation leads to the
result
\begin{eqnarray}
P(x,t) &=&\frac{N^{2}}{\sqrt{2\pi \Delta x^{2}(t)}}\left( \exp \left\{ -%
\frac{(x-\frac{d}{2})^{2}}{2\Delta x^{2}(t)}\right\} +\exp \left\{ -\frac{(x+%
\frac{d}{2})^{2}}{2\Delta x^{2}(t)}\right\} \right.   \nonumber \\
&&{}+\left. 2\exp \left\{ -\frac{x^{2}}{2\Delta x^{2}(t)}-\frac{d^{2}}{%
8\Delta x^{2}(t)}\right\} \cos \frac{\hbar tdx}{4m\sigma ^{2}\Delta x^{2}(t)}%
\right) .  \label{4s1}
\end{eqnarray}%
The first two terms are of the single-Gaussian form given by (\ref{4.4})
while the interference term is characterized by the cosine factor. The key
point to be made here is that this interference term persists for all time.
More generally, when either temperature or dissipative effects are present,
one measures the disappearance of the interference term i.e. the loss of
coherence (decoherence) by defining an attenuation coefficient $a(t)$, which
is the ratio of the factor multiplying the cosine to twice the geometric
mean of the first two terms. From an examination of Eq. (\ref{4s1}), we see
that $a(t)=1$, corresponding to the absence of decoherence.

To take into account the effect of finite temperature, we first form the
probability distribution (\ref{4s0}) with $\psi _{1}$ and $\psi _{2}$ given
by the form (\ref{4.3}) with $x_{0}=\pm \frac{d}{2}$ but keeping the terms
with $v_{0}$. It is a simple matter to see that the resulting probability
distribution can be obtained from (\ref{4s1}) with the replacement $%
x\rightarrow x-v_{0}t$. We then average this probability distribution over a
thermal distribution of initial velocities, as in (\ref{4.5}), to obtain the
probability distribution corresponding to a finite temperature $T$. After a
bit of algebra, the result can be written in the form%
\begin{eqnarray}
P_{{\rm T}}(x;t) &=&=\frac{N^{2}}{\sqrt{2\pi \Delta x_{{\rm T}}^{2}(t)}}%
\left( \exp \left\{ -\frac{(x-\frac{d}{2})^{2}}{2\Delta x_{{\rm T}}^{2}(t)}%
\right\} +\exp \left\{ -\frac{(x+\frac{d}{2})^{2}}{2\Delta x_{{\rm T}}^{2}(t)%
}\right\} \right.   \nonumber \\
&&\left. +2a(t)\exp \{-\frac{x^{2}}{2\Delta x_{{\rm T}}^{2}(t)}-\frac{d^{2}}{%
8\Delta x_{{\rm T}}^{2}(t)}\}\cos \frac{\hbar tdx}{4m\sigma ^{2}\Delta x_{%
{\rm T}}^{2}(t)}\right) ,  \label{4s1a}
\end{eqnarray}%
where the attenuation coefficient $a(t)$ is given by
\begin{eqnarray}
a(t) &=&\exp \{-\frac{\frac{kT}{m}t^{2}d^{2}}{8\sigma ^{2}\Delta x_{{\rm T}%
}^{2}(t)}\}  \nonumber \\
&=&\exp \left\{ -\frac{\frac{kT}{m}t^{2}d^{2}}{8\sigma ^{4}+8\sigma ^{2}%
\frac{kT}{m}t^{2}+\frac{2\hbar ^{2}t^{2}}{m^{2}}}\right\} .  \label{4s2}
\end{eqnarray}%
Once again, we see that $a(t)=1$ for $T=0$. However, for non-zero $T$ and
short times (characteristic of decoherence time scales), whereas the $t$
dependent terms in the denominator are negligible, the $t$ dependent terms
in the numerator remain and thus we obtain
\begin{equation}
a(t)\cong e^{-t^{2}/\tau _{d}^{2}},  \label{6.5}
\end{equation}%
where the decoherence time is
\begin{equation}
\tau _{d}={\frac{\sqrt{8}\sigma ^{2}}{\bar{v}d}},  \label{6.6}
\end{equation}%
and $\bar{v}=\sqrt{kT/m}$ is the mean thermal velocity. This is consistent
with the results obtained in Ref. 17-19 where we have found that the
dominant contribution to decoherence at high temperatures ($kT>>\hbar \gamma
$, where $\gamma $ is typical dissipative decay rate), is independent of
dissipation. However, for very low temperatures $T$, dissipation plays an
important role, in which case one must use sophisticated techniques from
non-equilibrium quantum statistical mechanics. \cite{ford1,ford3}

In order to see why decoherence is a short-time phenomenon,
consider as an example an electron at room temperature (300 K),
$\bar{v}=6.8\times{10}^{6}$ cm/s. Hence, if we take $d=1$ cm
and $\sigma=0.4 \AA$, then using Eq.~(\ref {6.6}), we obtain
$\tau_{d}=6.9\times{10}^{-24}$ s, which is orders of magnitude
smaller than typical $\gamma^{-1}$ values. For this reason it
is permissible to take $\gamma{t}<<1$ for calculations
involving the calculation of decoherence times and this is why
the simple derivation outlined above (which is solely within the
framework of elementary quantum mechanics and equilibrium
statistical mechanics) works.

\section{Conclusions}

Wave packet spreading is of fundamental interest in quantum mechanics. By
extending some of the usual methods, we have been led to many interesting
phenomena: temperature and squeezing effects on wave packet spreading with
applications to topical phenomena such as how the standard quantum limit may
be circumvented by squeezing and how the temperature dependence of the rate
of decoherence may be calculated in a simple manner. Concomitantly, we have
introduced simple tools and ideas that are at the frontiers of cutting-edge
research in quantum mechanics, such as Schr\"{o}dinger cat states,
decoherence, entanglement and the classical-quantum interface.

\acknowledgments We wish to thank the School of Theoretical Physics, Dublin
Institute for Advanced Studies, for their hospitality.

\end{document}